\documentclass[preprint]{raa}            % referee version: for submission

\usepackage{graphicx,times}             %for PS/EPS graphics inclusion, new
\usepackage{natbib}
\usepackage{amssymb,amsmath}
\usepackage{gensymb}
\bibpunct{(}{)}{;}{a}{}{,}

\definecolor{orange}{rgb}{1,0.5,0}

\usepackage[pagebackref=true]{hyperref}

\newcommand{\cm}{\ensuremath{\,{\rm cm}}}

\newcommand{\pc}{\ensuremath{\,{\rm pc}}}

\newcommand{\GHz}{\ensuremath{\, {\rm GHz}}}

\renewcommand{\Re}{\ensuremath{\,{\rm Re}}}

\begin{document}

%  \title{The FRB Searching Backend for the Tianlai Dish Pathfinder Array }
\title{A Fast Transient Backend to Detect FRBs with the Tianlai Dish Pathfinder Array}
%   \subtitle{I. Place Your Subtitle Here}

   \volnopage{Vol.0 (20xx) No.0, 000--000}      %%preserved for Editor. DOn't remove!
   \setcounter{page}{1}          %%starting page, preserved for Editor. DOn't remove!

   \author{Zijie Yu 
      \inst{1,2}
   \and Furen Deng
      \inst{1,2}
         \and Shijie Sun
      \inst{1}
   \and Chenhui Niu
      \inst{1}
   \and Jixia Li
      \inst{1}
   \and Fengquan Wu
      \inst{1}
    \and Wei-Yang Wang 
    \inst{3,4}
    \and Yougang Wang
    \inst{1}
    \and Hui Feng
    \inst{5,6}
    \and Lin Shu
    \inst{5,6}
    \and Jie Hao 
    \inst{5,6}
    \and Reza Ansari
    \inst{7}
    \and Albert Stebbins
    \inst{8}
   \and Xuelei Chen
      \inst{1,2,9,10*}
   }

   \institute{National Astronomical Observatories, Chinese Academy of Sciences,
             Beijing 100101, China; *{\it xuelei@cosmology.bao.ac.cn}\\
        \and
             University of Chinese Academy of Sciences Beijing 100049, China;\\
        \and 
            Department of Astronomy, Peking University, Beijing 100871, China;\\
        \and    
            Kavli Institute for Astronomy and Astrophysics, Peking University, Beijing 100871, China\\
        \and 
            Institute of Automation, Chinese Academy of Sciences, Beijing 100190, China;\\
        \and
            Guangdong Institute of Artificial Intelligence and Advanced Computing, Guangzhou 510535, China;\\
        \and
            Universit\'e Paris-Saclay, CNRS/IN2P3, IJCLab, 91405 Orsay, France;\\
        \and
            Fermi National Accelerator Laboratory, P.O. Box 500, Batavia IL 60510-5011, USA
        \and
            Department of Physics, College of Sciences, Northeastern University, Shenyang 110819, China\\
        \and 
            Center of High Energy Physics, Peking University, Beijing 100871, China
\vs\no
   {\small Received 20xx month day; accepted 20xx month day}}

\abstract{The Tianlai Dish Pathfinder array is a radio interferometer array consisting of 16 six meter dish antennas. The original digital backend integration time is at the seconds level, designed for HI intensity mapping experiment. A new digital backend with millisecond response is added to enable it to search for fast radio burst (FRB) during its observations. The design and calibration of  this backend, and the real time search pipeline for it are described in this paper. It is capable of forming 16 digital beams for each linear polarisation, covering an area of 19.6 square degrees. The search pipeline is capable of searching for, recording and classifying FRBs automatically in real time. In commissioning, we succeeded in capturing the signal pulses from the pulsars PSR B0329+54 and B2021+51. 
}

   \authorrunning{Zijie Yu et al. }            %author_head in even pages
   \titlerunning{The Tianlai Dish Pathfinder Array FRB Real Time Searching Backend}  % title_head in odd pages

   \maketitle

%% Note: In the following text body of your manuscript, please note several differences from
%%       other major journals:
%% (1) \subsection{Please Capitalize the First Letter of Each Notional Word in Subsection Title}
%% (2) Please Capitalize the First Letter of Each Notional Word in all tables' captions

%
%________________________________________________ sections below
%
\section{Introduction}           %% first-level sections will be auto-capitalized
\label{sec:introduction}

Since their  discovery\citep{Lorimer2007,Thornton2013}, fast radio bursts (FRBs) have been an intensely studied topic in radio astronomy.
The cosmological origins of them are established after FRB 121102 was localized to a host galaxy with $z=0.193$ \citep{Bassa17,Chatterjee17,Marcote17}.
The astrophysical origin of FRBs and radiation mechanism are still unsolved problems, but in the past few years significant progress have been made, with many more events observed by the various telescopes.
Up to now, over 800 fast radio bursts, including 25 repeaters, have been publicly announced \footnote{A catalogue is available at \url{https://www.herta-experiment.org/frbstats/}}.
The discovery of FRB 20200428 \citep{CHIME202005}, which is originated from the known galactic magnetar SGR J1935+2154, indicates that magnetars are probable sources for at least some FRBs (e.g., \citealt{Wang20}).

The FRB flash typically lasts a few milliseconds, so that the radio telescope must have a sub-millisecond sampling rate to detect the FRBs. Given the large amount of data received by the interferometer array, it is impractical to store all of the data at such high time resolution. Instead, the data within a short time frame is searched in real time for possible FRB, and the high time resolution data is stored for further analysis if a candidate is found, or else discarded if no FRB is found.

\begin{figure}
    \centering
    \includegraphics[width=0.5\textwidth]{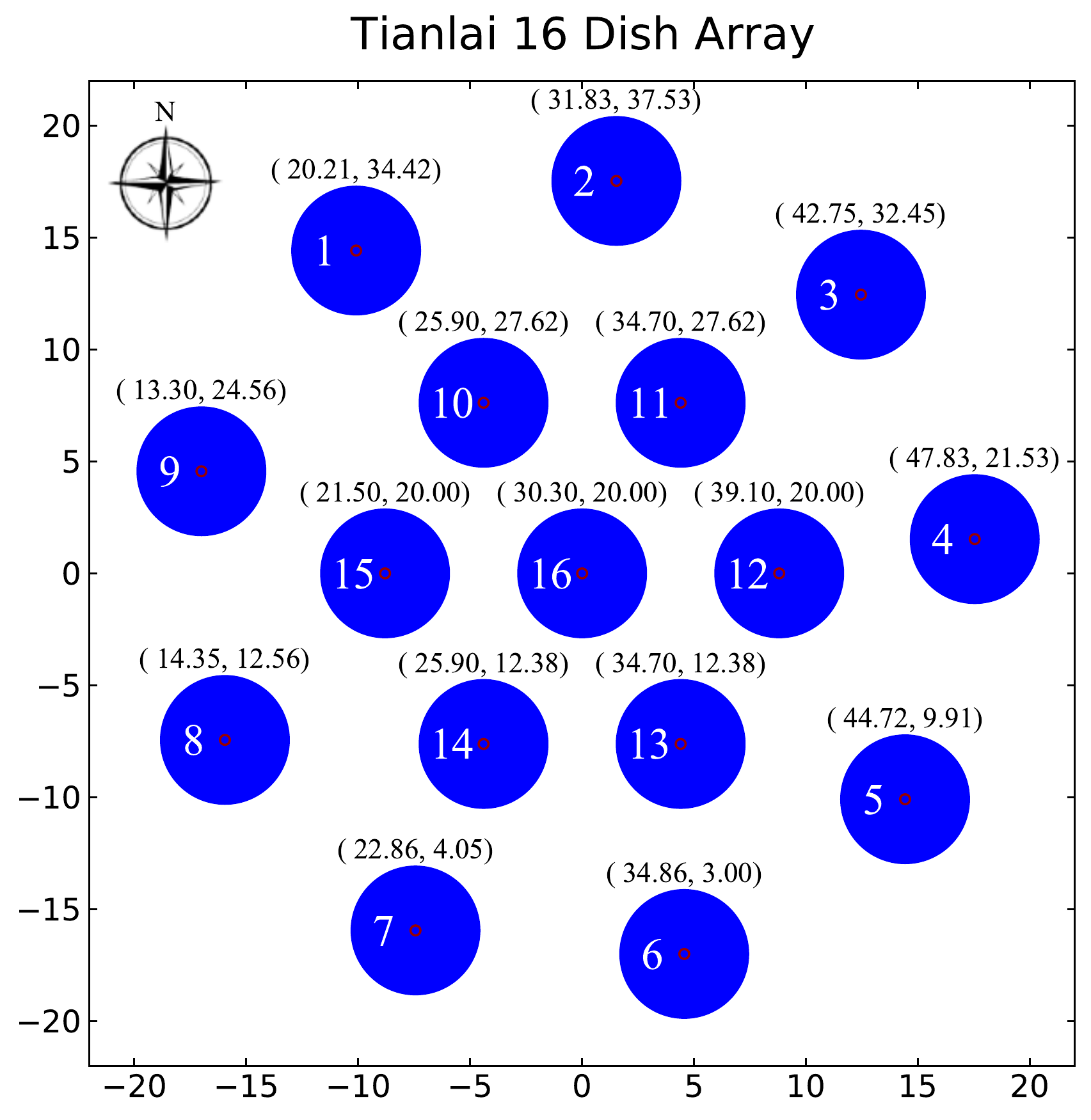}
    \caption{Tianlai dish antenna layout with antenna number and ground coordinates with respect to a reference point (unit:meter).}
    \label{fig:position}
\end{figure}

This paper describes the FRB digital backend of the Tianlai dish pathfinder array. The Tianlai experiment is dedicated to test the key technologies for 21-cm intensity mapping method \citep{Chen2012,Xu_2014}, which consists a cylinder pathfinder array\citep{Li:2020ast,Zhang_2016,Sun_2022,LI_2021} and a dish pathfinder array \citep{Wu2021dish,PAON4_Zhang_2016,Das:2018nwp,Phan_2022}. It  consists of 16 on-axis dish antennas of 6-meter aperture, equipped with dual linear polarization receivers, its basic parameters are listed in Table \ref{tab:properties}, and the array configuration is shown in Fig.~\ref{fig:position}. The original Tianlai correlator is only designed to output visibilities with a $\sim 1$ second time sampling, which is then  processed offline for sky imaging \citep{ZUO2021100439}. However,  with a field of view (FoV) of $\sim$ 19.6 square degrees, the array is potentially capable of discovering large numbers of FRBs while doing its HI survey \citep{Perdereau:2022ksl}, if it is equipped with digital backend which has millisecond time resolution to perform FRB searches.  

\begin{table}
    \centering
    \caption{Basic properties of Tianlai Dish Pathfinder Array}
    \begin{tabular}{c c}
    \hline 
         Antenna mount & Alt-Az pedestal\\
         Number of Antenna  & 16            \\
         Dish diameter & 6m              \\
         f/D & 0.37                      \\
         Average SEFD & 14.15 kJy        \\
         Latitude & 44.15\degree N       \\
         Longitude & 91.80\degree E      \\
         Observing frequency & 685-810MHz          \\
         
    \hline
    \end{tabular}
    \label{tab:properties}
\end{table}

Below, we first present the operation principle and design of the Tianlai Dish Array FRB backend, then  describe our commissioning experiment and test results, followed by an estimate of its sensitivity and the expected number of FRBs. Finally we summarize the results . For the antennas in the Tianlai dish array, we use labels explained in \citet{Wu2021dish} and also shown in Fig.~\ref{fig:position}.

\section{System Design}
\label{sec:System Design}
The FRBs occur randomly in any direction of the sky, so an blind FRB survey can be conducted by pointing the telescope in any direction and searching for burst events in the output. The FRB signal is dispersed by the interstellar medium, so that the higher frequency part arrives slightly earlier, with the time delay (in units of seconds) with respect to $\nu\rightarrow \infty$ given by 
\begin{equation}
\Delta t \approx 4.148 \times 10^{-3} \frac{\rm DM}{\pc \cm^{-3}} \left(\frac{\nu}{\GHz}\right)^{-2} 
\end{equation}
where ${\rm DM}$ is the dispersion measure along the propagation path. The dispersion phenomenon can be used as a character to distinguish the FRB or pulsar signal from the radio frequency interference (RFI), which occurs much more frequently.

The conventional correlator for the Tianlai array produces interferometric visibilities, i.e. the short time integration of the cross-correlation of voltages of different receiver pairs, and the sky image is synthesized from such visibilities, which corresponds approximately to Fourier components of the sky radiation intensity \citep{Thompson_book}. However, there are a large number of visibilities (528 including auto-correlations), the individual sensitivity to a spatially localized source such as an FRB is not high.  To facilitate the search of FRB, we adopt a beam-forming approach for the FRB search engine. We digitally form multiple beams sensitive to different sky directions, the output data is then de-dispersed with a range of dispersion measure (DM) values and searched for pulses above the noise. This is realized with an independent digital system, which shares the analog frontend hardware of the radio array with the original correlator, but operates independently. Below we describe this digital system.

\subsection{Beam Forming principle}
 
 The voltage output from antenna unit $a$ is:
\begin{align}\label{eq:voltage}
    \mathcal{E}_a = |g_a| e^{j \phi_a}\int e^{-2\pi j\bm{n}\cdot \bm{u}_a} A_i(\bm{n}) I(\bm{n}) d^2\bm{n} + \eta_a, 
\end{align}
where $A_a(\bm{n})$ describes the voltage response for antenna $a$, $\bm{u}_a$ is the position vector of the antenna in unit of wavelength, the integration is over sky directions, and $g_a= |g_a| e^{j \phi_a}$ is the complex gain of the instrument for unit $a$, $\eta_a$ represents the noise for that channel.  For the Tianlai array, the instrument phase $\phi_a$ comes mainly from the optical cable, though the feed and amplifier/bandpass filter may also contribute. We can form a digital beam from the outputs of the array by adding the voltages with different complex weights, 
\begin{equation}
S (\bm{k}) = \sum_a w_a(\bm{k})  \mathcal{E}_a
\label{eq:sum}
\end{equation}
where $w_a(\bm{k})$ denotes the complex weight of antenna $a$ for beam direction $\bm{k}$, which we can set as 
\begin{equation}
w_a(\bm{k})= \frac{1}{|g_a|} e^{-j \phi_a} e^{2\pi j \bm{k} \cdot \bm{u}_a},
\label{eq:weight}
\end{equation}
so that ideally, 
 \begin{equation}
 S (\bm{k}) = \sum_a \int e^{-2\pi j\bm{(n-k)}\cdot \bm{u}_a} A_i(\bm{n}) I(\bm{n}) d^2\bm{n}.
 \label{eq:beamforming}
 \end{equation}
In this integral, the phases of all antennas are equal for the direction $\bm{n}=\bm{k}$, so that their voltages add coherently, while for other directions the terms in the sum would be out of phase, so this output is most sensitive to the direction $\bm{k}$.  If a burst occurs in the direction $\bm{q}$ at some moment, and assume that sky in other directions remains nearly constant at this time, then the received signal is given by 
\begin{equation}
  S= S_0 + \Delta S  
\end{equation}
where 
\begin{equation}
\Delta S =\sum_a e^{-2\pi j(\bm{q}-\bm{k})\cdot \bm{u}_a} A_i(\bm{q}) I (\bm{q}),    
\label{eq:DeltaS}
\end{equation} 
and $S_0$ is the S value without burst.

\begin{figure}
    \centering
    \includegraphics[width=0.49\textwidth]{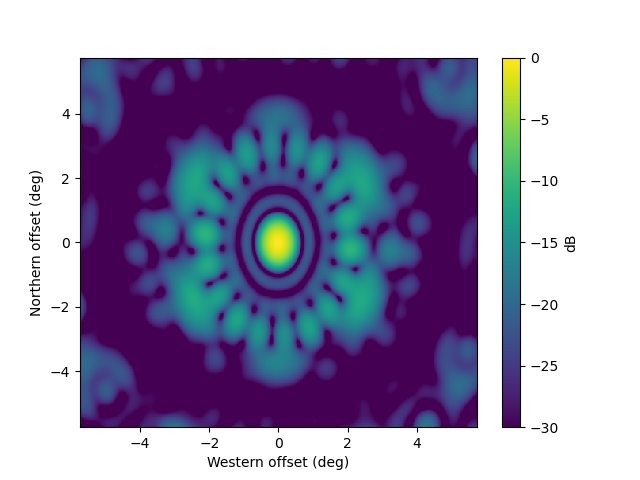}
    \includegraphics[width=0.49\textwidth]{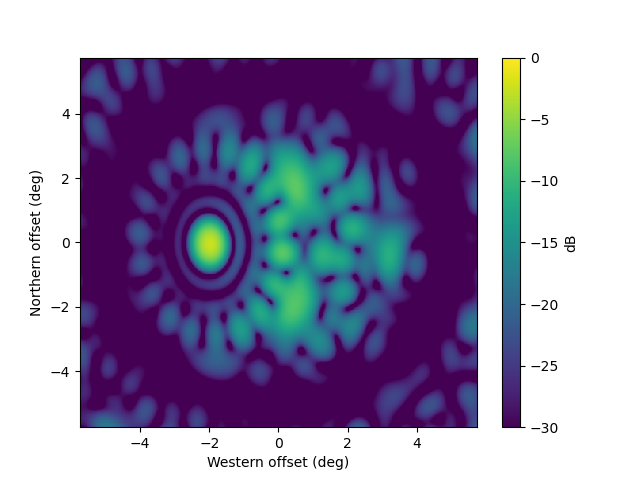}
    \caption{ The profile for digitally formed beam in dB at 750 MHz. We assume all the primary beams are identical and pointed to the North Celestial Pole (NCP). Left: the digital beam centered at NCP. Right: the digital beam centered at the same elevation, but $2^\circ$ East of NCP. The amplitudes of beams are normalized so that the beam centered at NCP has maximum amplitude of 0 dB. }
    \label{fig:beam}
\end{figure}

The primary beam of the Tianlai dish has been measured with unmanned air vehicle \citep{Zhang_2021}. Here 
we simply model the primary beam as
\begin{equation}
A(\theta) = A_0 \mathrm{sinc}\left(\frac{\theta}{2\sigma}\right)    
\label{eq:primarybeam}    
\end{equation}
where $\theta$ is the angle between the direction of interest to the direction of antenna axis and $\sigma\approx 0.79^\circ$ corresponding to an full width at half maximum (FWHM) $\approx 4.4^\circ$.  The digitally formed beam can point to any direction, but for different directions the beam shape would be somewhat different, as the antennas are fixed in position and the projected baselines are different for different directions.

In Fig.\ref{fig:beam} we show the synthesis beam centered at the North Celestial Pole (NCP) (Left), and the beam centered at a point at the same 
elevation of NCP but $2^\circ$ to its East (Right). The FWHM of the main lobe of the synthesis beam is about $0.6^\circ$ for the east-west direction and $0.8^\circ$ for the north-south direction.  There are also sidelobes which extends over the primary beam. Note that the profile of both the primary beam and the digitally formed beam vary with frequency, but for the current Tianlai arrays which have a fractional bandwidth of $\sim 13\%$ at 750 MHz the variation is moderate.

\subsection{The system setup}

\begin{figure}
\centering
\includegraphics[width=0.95\textwidth]{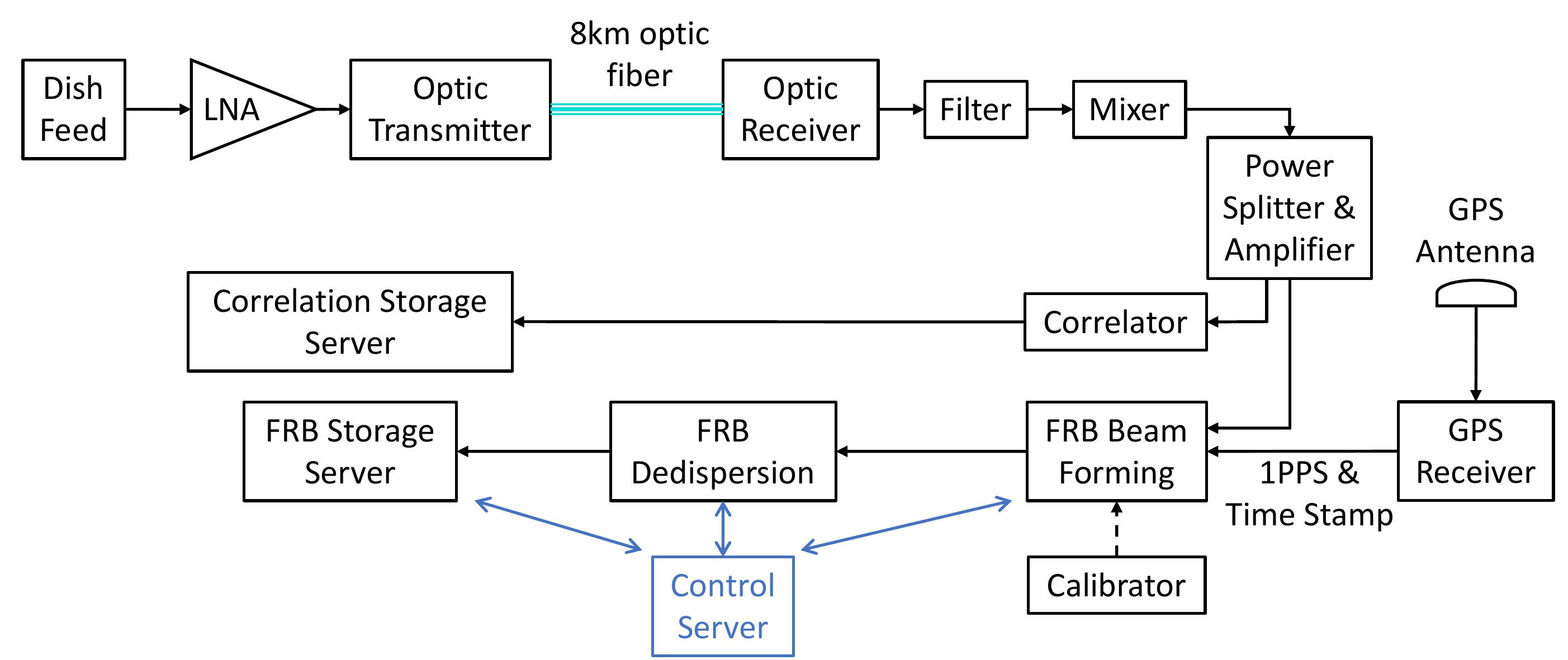}
\caption{Schematic of the Tianlai FRB-search and correlation system.}
\label{fig:whole_schematic}
\end{figure}

The FRB system shares the same analog hardware system with the original Tianlai dish correlator system, as shown in Fig.~\ref{fig:whole_schematic}. A detailed description of the analog system can be found in \citet{Wu2021dish}, here we give a brief summary. 
The Tianlai dish array consists of 16 steerable parabolic reflector antennas. Each antenna is equipped with a dual, linear-polarization feed, thus 32 polarized radio signals are collected in total. The signals are amplified by the low noise amplifiers (LNA), which are installed on the backside of the feed unit, and transmitted by a 15-meter long cable down to the optic transmitters below the reflectors. In the Tianlai experiment, the station house where the signal chain and digital facilities are running are located in a village which is 6 km away from the antenna area. The radio signals are converted to optical signals at the antenna site and communicated through optical fibers to the station house, where they are converted back to radio frequency (RF). Then, the RF signals are bandpass filtered for a selected 100 MHz bandwidth (currently in the 700--800 MHz, but maybe shifted to other frequency bands). The signals are then down-converted to the intermediate frequency (IF) with a range of 135--235MHz via a mixer. The IF signals are further amplified. Up to this point, the FRB system and the original correlator system share the same signal chain hardware. 

Each of the IF signal is then split into two channels by a power splitter. The signal is amplified with an additional amplifier to its original magnitude to compensate for the power loss. One of signal is fed to the original digital correlator, while the other is fed to the FRB searching system. 
A schematic of the whole Tianlai Dish Array  system is shown in Fig.~\ref{fig:whole_schematic}.

The hardware of the FRB digital backend consists of FPGA-based beamformers and GPU-based de-dispersion servers. The beamformer digitize the signal, and form multiple beams according to Eq.~(\ref{eq:sum})-(\ref{eq:DeltaS}). The output of the beamformer is send to the de-dispersion server, which de-disperse the signal with a range of DM values and check for possible FRB candidates. 
At present, the Tianlai Dish Array beamformer consists of two FPGA boards, each processes the input from one linear polarization of the feeds, and correspondingly there are also two GPU servers. The beam former time resolution is up to 98.304 $\mu s$ with 1024 frequency channels and 122.07 kHz channel width. The FPGA-based beamformer and the real time processing of the beamformed data by the de-dispspersion system are described in the next two subsections. 

\subsection{The FPGA Beamformer}

As shown in Fig.\ref{fig:whole_schematic}, the FPGA beamformer for the Tianlai dish array consists of two CASPER SNAP2\footnote{https://casper.astro.berkeley.edu/wiki/Hardware} boards \footnote{Earlier we have also tested a system with ROACH boards \citep{2019RAA....19..102N}, where a detailed description of the design of the hardware architecture and data flow was given. Although in this implementation a different board is used, the data flow structure is very similar, and we refer the reader to that paper for more details.}, each of them processes data from one of the linear polarizations. Each SNAP2 board have two analog-to-digital converter (ADC) boards, and each ADC converts analog input signal from 8 units into digital data. Each SNAP2 board processes the raw data from both ADCs, and output the beamformed intensity data to a de-dispersing GPU server through four 10GbE network interface cards (NICs).  For system testing, the ADCs are also capable of returning the real-time data directly, bypassing the data processing on the SNAP2 board. 

\begin{figure}
    \centering
    \includegraphics[width =0.8\textwidth]{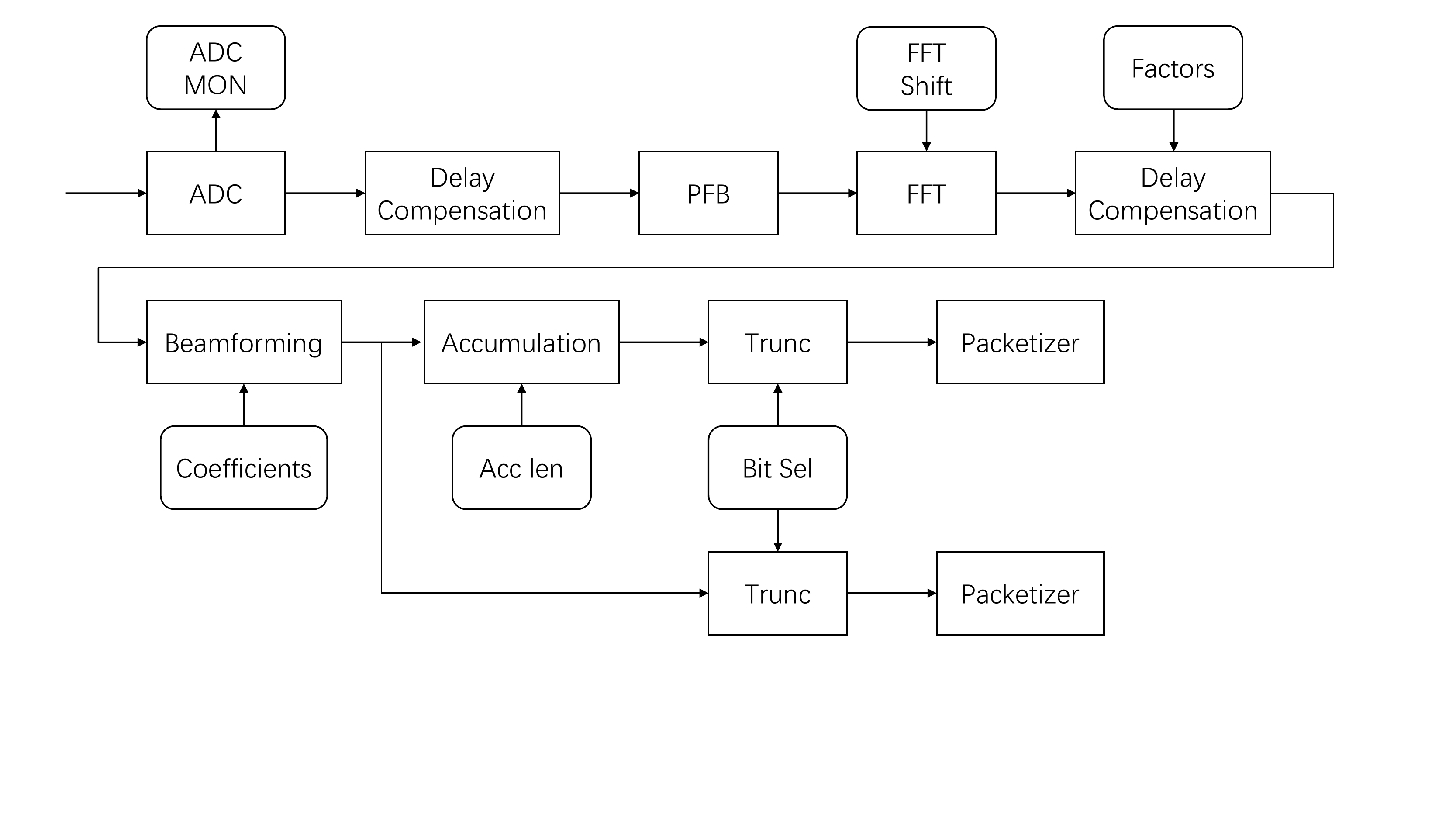}
    \caption{The workflow of the Tianlai dish array beamformer.}
    \label{fig:workflowbeamformer}
\end{figure}

The workflow of the beamformer is shown in Fig.\ref{fig:workflowbeamformer}. The raw data collected by the ADCs needs to be calibrated for delay and phase compensation, due to the different cable lengths and the change of geometric delay when the antennas are pointing to different directions. For the Tianlai Dish Pathfinder Array, the maximum signal delay between two feeds is about 50 ns. 
This delay is compensated in two steps. The first step deals with the major part of the delay up to multiples of the sampling period which is 4ns. This is done by shifting the data block. Then, in the second step, after polyphase filtering and fast fourier transforming the data to frequency domain, the residual of the delay is compensated by a phase factor, which is given by the sum of geometric delay which is computed for each pointing direction, and the instrument delay which is determined by calibration with astronomical source. Compensation coefficients are then uploaded to the beamformer to fulfill Eq.(\ref{eq:sum}).

The data of each beam is accumulated with a preset integration time, the minimum is 0.1 ms. The accumulated data is truncated into 8 bit integers, and sent to GPU server in UDP packets through the 10GbE NICs. The raw data flow is also monitored for anomalous values, which will raise alerts for possible system problems.

\subsection{Data Real-Time Processing}
The stream data of the digitally formed beams are further processed on the two GPU servers, each equipped with an Intel Xeon E5-2690 V4 CPU, 128GB RAM and two NVIDIA GTX 1080Ti GPUs. The pipeline consists of five parts. The data packets sent by the FPGAs are received by a running High Availability Shared Pipeline Engine (HASHPIPE) \footnote{https://casper.astro.berkeley.edu/wiki/HASHPIPE} program, which assembles the packets to a form appropriate for processing on the GPU. De-dispersion and FRB candidates searching are then done on the GPU. In principle, the incoherent de-dispersion and FRB can be done by any of the algorithm and software available, and we are working on the development of an algorithm based on the Hough transform \citep{Zuo_2020}, but at present it is done by the software HEIMDALL\footnote{https://sourceforge.net/projects/heimdall-astro/} \citep{Barsdell2012}. The candidates found by HEIMDALL are screened to remove obvious RFIs, the program for this is described later.  The remaining candidates are classified as possible FRB candidates or RFIs by the deep learning program FETCH\footnote{https://github.com/devanshkv/fetch} \citep{Agarwal2020} program. In the end, the FRB candidates are plotted and their information will be saved to enable human inspection. The data processing are done with multiple threads running on the server, and flag files are used for communication between the threads.

\begin{figure}
    \centering
    \includegraphics[width=0.95\textwidth]{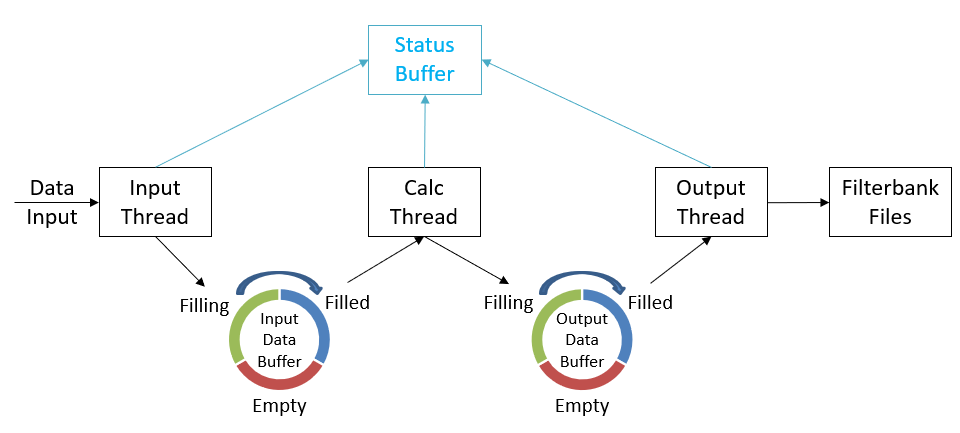}
    \caption{The HASHPIPE  pipeline thread. 
    }
    \label{fig:hashpipe}
\end{figure}

This pipeline is capable of processing beamformed data  automatically and in quasi-real time. Each GPU server is connected to one FPGA beamformer with four Gbps (Gigabit SFP+) cables, and each cable transports data from four beams. We create four HASHPIPE threads, each processing data from one beam. A HASHPIPE thread has two data buffers and three child processes, as Fig.\ref{fig:hashpipe} shows. Each data buffer contains 3 looped sub-buffers. Input thread reads data packets from network sockets and writes them into one of the sub-buffers. When a sub-buffer is full, input thread will writes the data into another sub-buffer and calc thread will start processing the filled sub-buffer. Calc thread re-arranges the data structure and write them into output data buffer so that output thread is able to write the data in the {\tt Filterbank} format files for further process by HEIMDALL. To improve the pipeline's I/O performance, files are placed in RAM instead of disk. Each {\tt Filterbank} file contains 201.3 seconds of data. In other words, we need to complete processing existing data within 201.3 seconds before new data arrives in order to build a quasi-real-time processing pipeline. The total data stream is 159.1 MB/s.

\section{Experiment}
\subsection{Calibration}
\label{sec:calibration}

\begin{figure}
    \centering
    \includegraphics[width=1.0\textwidth]{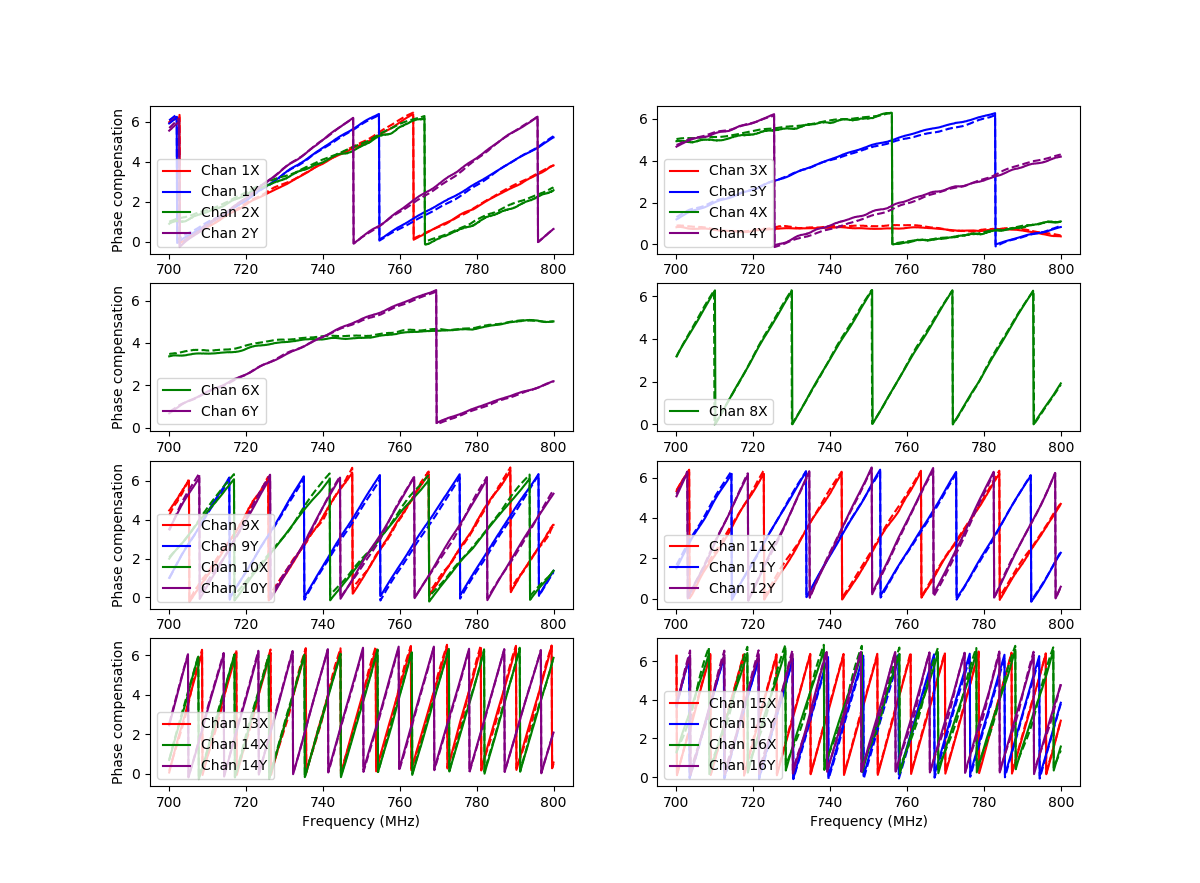}
    \caption{Result of $\phi_{ab}$ for two sequential calibrations. The solid line shows the first calibration and dashed line shows the second. Each figure show 4 feed channels, malfunctioning ones are not shown. The reference channels 5X and 5Y channels are not shown.}
    \label{fig:calibration_validate}
\end{figure}

\begin{figure}
    \centering
    \includegraphics[width=0.8\textwidth]{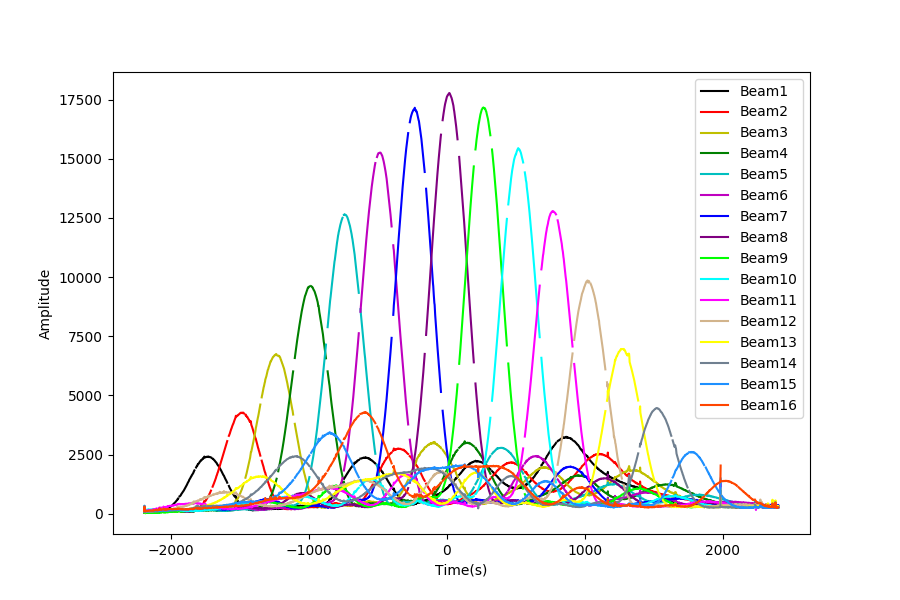}
    \caption{The output amplitude of the 16 beams as a function of time. We set the beams along the track of  Cas A with a spacing of $0.5^\circ$. }
    \label{fig:beam_pointing_test}
\end{figure}

\begin{figure}
    \centering
    \includegraphics[width=0.8\textwidth]{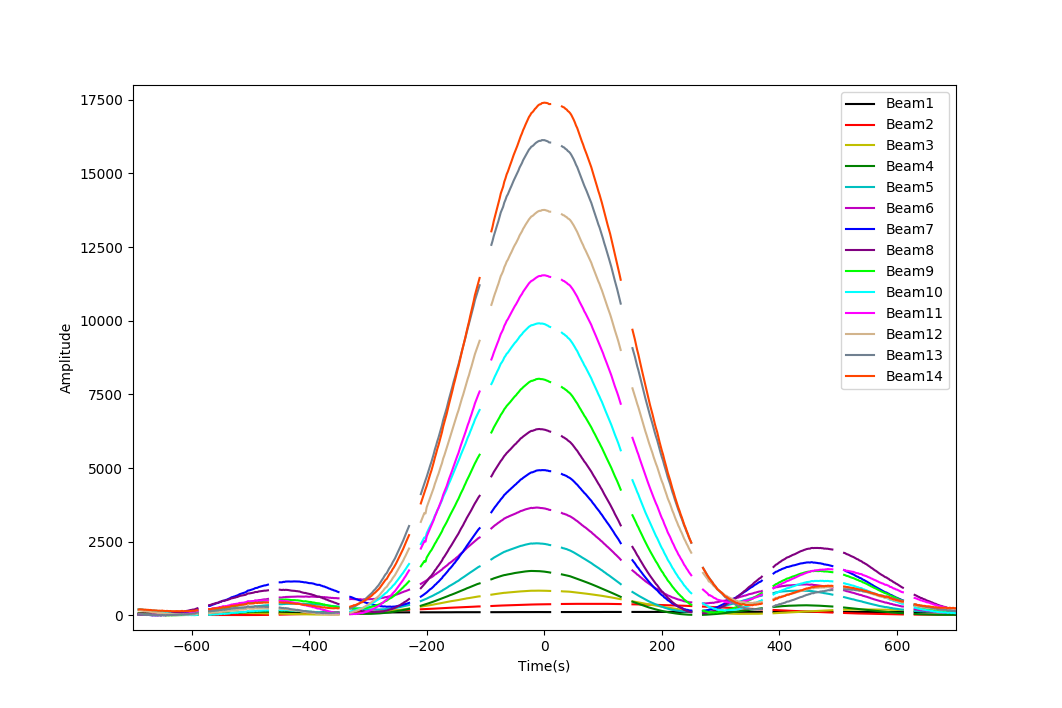}
    \caption{Beam amplitude vs. Time, where there are 14 beams synthesized from signal from 1-14 antennas, respectively.}
    \label{fig:beam_amp_test}
\end{figure}

The complex gain $g_a$  needs to be determined by a calibration procedure. One possibility is to use the original correlator to do this calibration \citep{zuo2019eigenvector}, but the separate circuit for the original correlator and the FRB beam former may introduce some differences. We have therefore adopted an self-calibration approach based entirely on the FRB beam former.

The absolute value of each complex gain is first adjusted so that the amplitude of the readout for each feed is approximately equal (within 1 dB). The instrument phase is then calibrated by observing the transit of a bright point source, during which we make special beams used for calibration. In each such special beam, all weights are set to be zero except for a reference antenna and the antennas to be calibrated.  In the ground coordinate system, the antennas are fixed while the source moves with the rotation of celestial sphere, so the output of this beam for a pair of antennas $a,b$ is 
\begin{align}\label{eq:cal_i}
    I_{ab}(t) &= |g_a|^2 |A_a(\bm{n})|^2 + |g_b|^2 |A_b(\bm{n})|^2 + |g_a| |g_b| \left[A_a(\bm{n})A_b^*(\bm{n}) e^{-j[2\pi(\bm{n}-\bm{k})\cdot \bm{u}_{ab}+ (\phi_a - \phi_b)]}+\mathrm{c.c.}\right],
\end{align}
where c.c. denotes complex conjugate, $\bm{u}_{ab}=\bm{u}_b - \bm{u}_a$, and $\bm{n}(t)$ is the position of point source at time $t$ which is known (we have omitted to write out the $t$), and $\bm{k}$ is the direction of the primary beam center. Denote $\varphi_{ab}(t) = 2\pi\bm{n}(t)\cdot \bm{u}_{ab}$, we model the primary beam $A$ as a sinc function, fitting Eq.(\ref{eq:cal_i}) by:
\begin{align}
    \Re I_{ab}(t) = A^2\left[\cos(\varphi_{ab}(t) - \varphi_{ab}(t_{\rm transit}) + \phi_{ab}) + C^2\right]{\rm sinc}^2\left(\frac{\varphi_{ab}(t) - \varphi_{ab}(t_{\rm transit})}{2\sigma}\right),
    \label{eq:fit}
\end{align}
where $t_{\rm transit}$ is the time for the source transit (passing the beam center). The instrument phase difference 
$\phi_{ab}=\phi_a-\phi_b$ 
is then determined, assuming it to be constant during the calibration process.

The beamformer allows us to simultaneously form up to 16 beams for each polarization. From each of these the instrument phase difference of that pair can be solved by the transit observation. If the set of pairs include all 16 antennas, then in principle the instrument phase of all 16 antennas can be solved, up to an overall arbitrary phase offset, which can be set to zero for the reference antenna. However, we need to use baselines which have large variation of geometric phase during the transit to obtain good solution for $\phi_{ab}$. Due to the circular geometry of the Tianlai array, this is not possible for a single reference antenna. Instead, we choose two reference antennas during each calibration. For example, in one calibration of Cas A transit, we can 
choose antenna 1 and 5 which are located at opposite sides of the array as reference antennas.
The remaining antennas are either paired with antenna 1 or antenna 5, whichever allows a longer projected baseline for the transit position. The instrument phases of the antennas with respect to either antenna 1 and antenna 5 are then obtained. The antenna 1 and 5 are also paired, so that the instrument phase difference between the two is obtained. Then the phase of everyone antenna of the array can be solved with respect to the designated antenna. 

In Fig.\ref{fig:calibration_validate}, we show the instrument phases with respect to antenna 5 for each feed in two sequential calibrations separated by an hour. To check the reliability of the calibration, we have used two different set of references. In the first observation, antenna 3 and 14 are used as reference antennas, while in the second one antenna 1 and 5 are used.  The phase obtained from the two calibrations agree very well, showing that our results are consistent for whichever pair of reference antennas, and the instrument remains very stable during night time. 

\begin{table}
    \centering
    \caption{Beam output vs. number of antenna in formed beam. }
    \begin{tabular}{c c c c}
    \hline 
         Antenna Number & Prediction  & Measurement & Efficiency \\
         \hline                      %peak  
         1 & $1$  & 1 & 1\\ %121.4
         2 & $4$ & 3.23 & 0.81\\ %393.3
         3 & $9$ & 6.91 & 0.77\\ %839.1
         4 & $16$ & 12.4 & 0.78\\ %1507
         5 & $25$ & 20.1 & 0.80\\%2446
         6 & $36$ & 30.1 & 0.84\\ %3659
         7 & $49$ & 40.6 & 0.83\\%4930
         8 & $64$ & 52.1 & 0.81\\%6325
         9 & $81$ & 66.1 & 0.82\\%8028
         10 & $100$ & 81.6 & 0.82\\%9911
         11 & $121$ & 95.1 & 0.79\\%11541
         12 & $144$ & 113 & 0.78\\%13760
         13 & $169$ & 133 & 0.79\\%16130
         14 & $196$ & 142 & 0.72\\%16130
%         15 & $225V^{2}$ & 225 & 0.35\\%17270
%         16 & $256V^{2}$ & 256 & 0.3\\%17400
    \hline
    \end{tabular}
    \label{tab:beam_amp_test}
\end{table}

With the instrument phase determined, we can now form beams using all antennas and make observation. We test our beam pointing accuracy by making   beams along the track of a bright source, with a spacing of  $0.5^\circ$.  We expect to see the output amplitude of each formed beam reaches peak one by one as the order of beam number. The result is shown in Fig.\ref{fig:beam_pointing_test} as anticipated. 

If the complex gains of all antennas are perfectly accurate, the output of all antennas will add coherently, then for $n$ antennas, the output for a source at the beam center should $\propto n^2$. In reality, there will be errors in the calibration of the amplitudes and phases of the antennas, then when we form the beams, the output would be less than that of $n^2$.  We test the efficiency of beamforming by observing the strong radio source Cas A. We first use the transit 
of Cas A to calibrate the complex gains. Then, we re-pointing our antennas so that the Cas A transit them again. During this second transit, we form the beams 
as follows: all beams are pointing to the same direction, beam $\#1$ has non-zero weight for antenna 1 only; beam $\#2$ has  non-zero weight for antenna 1 and 2; beam $\#3$ has non-zero weight for antennas 1,2,3; and so on. As we have two malfunctioned antennas during the test, we formed up to 14 beams, with beam $\#14$ include all 14 working antennas. The transit curves are shown in Fig.\ref{fig:beam_amp_test}. The expected output, the measured output, and the
beamforming efficiency (defined as measured/expect) are given in  Table \ref{tab:beam_amp_test}. The typical efficiency achieved is about 0.8.

\subsection{De-dispersion}
We use HEIMDALL, a GPU accelerated radio transient detection pipeline, to search for single pulses in our data. For $m$ DM trials and $n$ boxcar width trials, HEIMDALL creates $m\times n$ de-dispersed and downsampled time series to cover parameter space. For each time series, the median value of the data is subtracted and the data is divided by the standard deviation so as to convert amplitude data into signal-to-noise ratio (SNR) data. Samples in the SNR data that exceed a preset threshold, usually 6, are marked as ``giants". Giants from different beams are checked for coincidence, those with the same time and DM are joined as one candidate. Finally, HEIMDALL outputs all candidates found in a candidate file.  

The HEIMDALL parameters we set are listed in table \ref{tab:heimdall_config}. For candidates signal-to-noise ratio threshold, we use 6, the default setting. DM search range is set to be 10 to 2000. Most RFIs can be excluded by the lower DM limit, as they do not exhibit dispersion. The upper limit is set to limit the required computation. The Tianlai dish array is a very small one, with quite limited sensitivity, so it is not very likely to detect very distant FRBs with large dispersion. The SNR loss tolerance is the maximum SNR loss allowed between DM trails. Under our setting, HEIMDALL has a total of 874 DM trials for each data set. The maximum boxcar width sets the upper limit of pulse width that HEIMDALL searches for. It is set to be 128 samples, which means we search pulse width up to 25.2 milliseconds. Because strong RFIs may create huge amount of candidates and lead to pipeline failure, a maximum giant rate is set to restrict the maximum candidates per minute that HEIMDALL report. If the number of candidates exceeds the threshold, HEIMDALL will abort further searching and report the found candidates.

\begin{table}
    \centering
    \caption{HEIMDALL configuration}
    \begin{tabular}{c c}
    \hline
        SNR threshold & 6 \\
        DM range & 10 - 2000 \\
        SNR loss tolerance & 1.5 \\
        DM trials & 866 \\
        Max boxcar width & 128 \\
        Max giant rate & 2000 per minute \\
    \hline
    \end{tabular}
    \label{tab:heimdall_config}
\end{table}

\subsection{Candidates filtering and classification}
\begin{figure}
    \centering
    \includegraphics[width=0.7\textwidth]{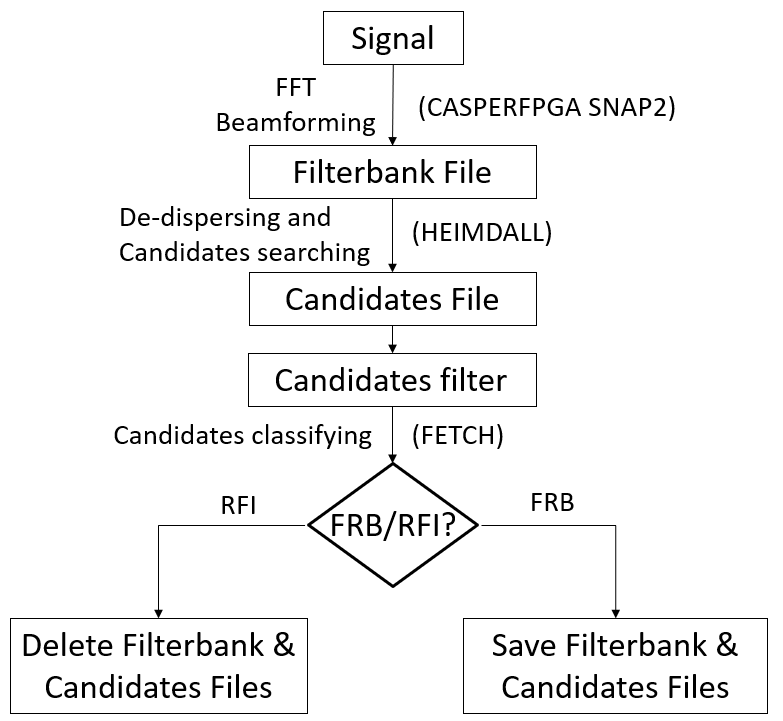}
    \caption{Overall workflow of our pipeline}
    \label{fig:pipeline}
\end{figure}

With these settings, HEIMDALL reports thousands of candidates with $S/N > 6$ per minute. We further filter these candidates out to remove obvious RFIs. The pipeline is shown in Fig.~\ref{fig:pipeline}.

Candidates with total $S/N<7$ are removed, and those with $\mathrm{DM} < 20 \pc\cm^{-2}$ are considered to be RFI. The DM criteria here is set at a higher value than the one set for HEIMDALL, in order to remove very strong and broad RFIs, which may still be strong enough to become candidates when the data is de-dedispersed to DM=10.  

As we form our beams in different directions while searching for FRBs, candidates found in all beams at the same time is most likely an  artificial radio frequency interference, which is either in the near field of the array, or too strong. Such events are filtered out. If all candidates within a time frame are rejected by the filter program, the data files in the RAM are removed. Data files with candidates passing through the filter are saved to the disk. 

\begin{figure}
    \centering
    \includegraphics[width=0.95\textwidth]{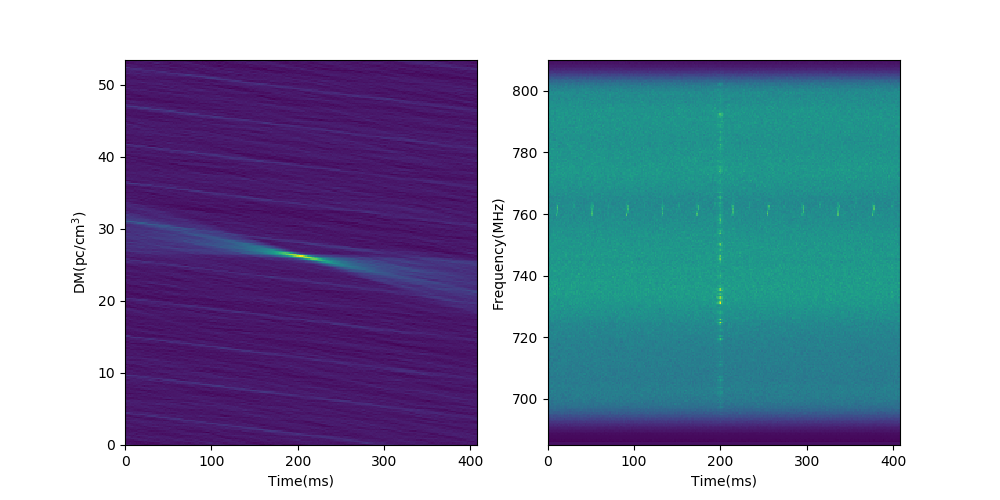}
    \caption{The FETCH candidate plot of a pulse from PSR B0329+54. Left: time-DM plot. Right: time-frequency (waterfall) plot. The band at 760 MHz is RFI.}
    \label{fig:fetch_candidate}
\end{figure}

We next use the Fast Extragalactic Transient Candidate Hunter (FETCH) \citep{Agarwal2020}, a deep-learning based classifier, to classify the remaining candidates. FETCH takes the SNR of the candidate, the starting time of the burst, the DM value and the box-car filter width reported by the HEIMDALL program as its input. It plots the output amplitude distribution in the time-DM and time-frequency figure for each candidate. An example is shown in Figure \ref{fig:fetch_candidate} for the pulsar test observation (described below). A pre-trained deep learning model takes these two figures as input and output the possibility of the candidates being a genuine astronomical source. Candidates with possibility greater than 50\% will be regarded as plausible FRBs. Data files with no plausible candidates are removed. 

\begin{figure}
    \centering
    \includegraphics[width=0.95\textwidth]{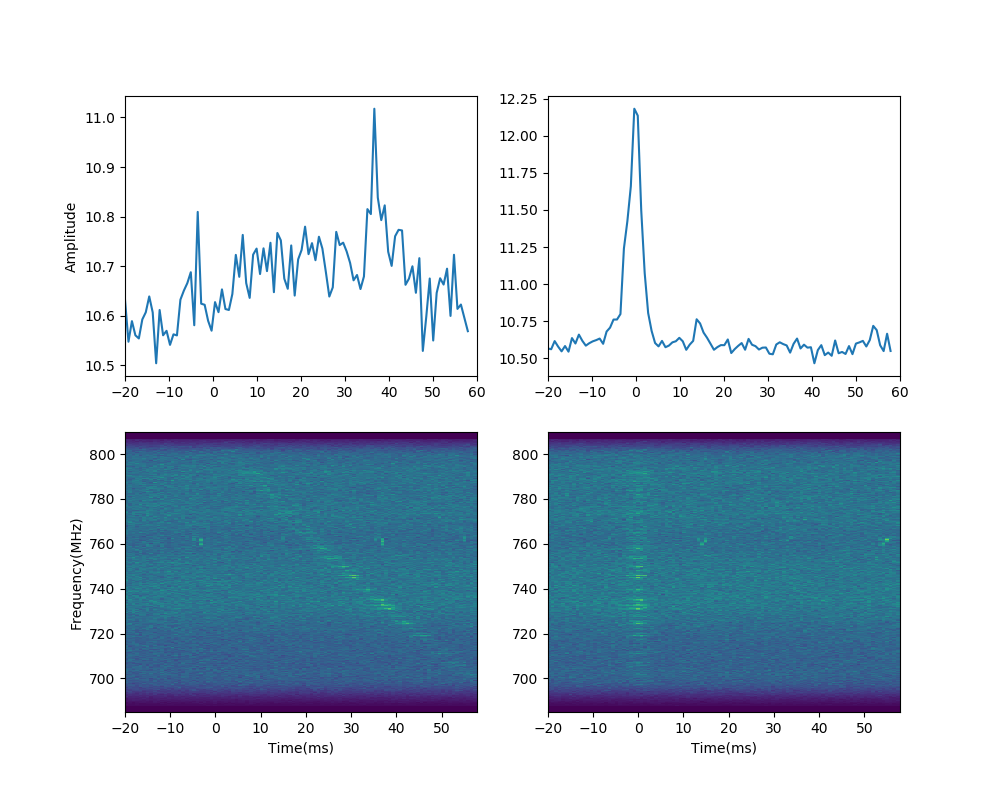}
    \caption{The dynamic spectrum of the pulse in Fig.\ref{fig:fetch_candidate}. Left:  before dedispersion. Right: after de-dispersion.}
    \label{fig:candidate_picture_sample}
\end{figure}

\subsection{Pulse searching test}
\label{sec:pulse_searching}

We tested our system with known pulsars. The pointing of our beams is put to test by observing the brighter pulsar PSR B0329+54. We evenly spaced our 16 beams along the pulsar's transit track with a separation of $0.5^\circ$, and then carried out an observation. The data is run through the pipeline, and the pulses from the pulsar are successfully detected. In Fig.~\ref{fig:candidate_picture_sample}, we show the dynamic spectrum of a pulse before (left) and after de-dispersion. 

\begin{figure}
    \centering
    \includegraphics[width=0.8\textwidth]{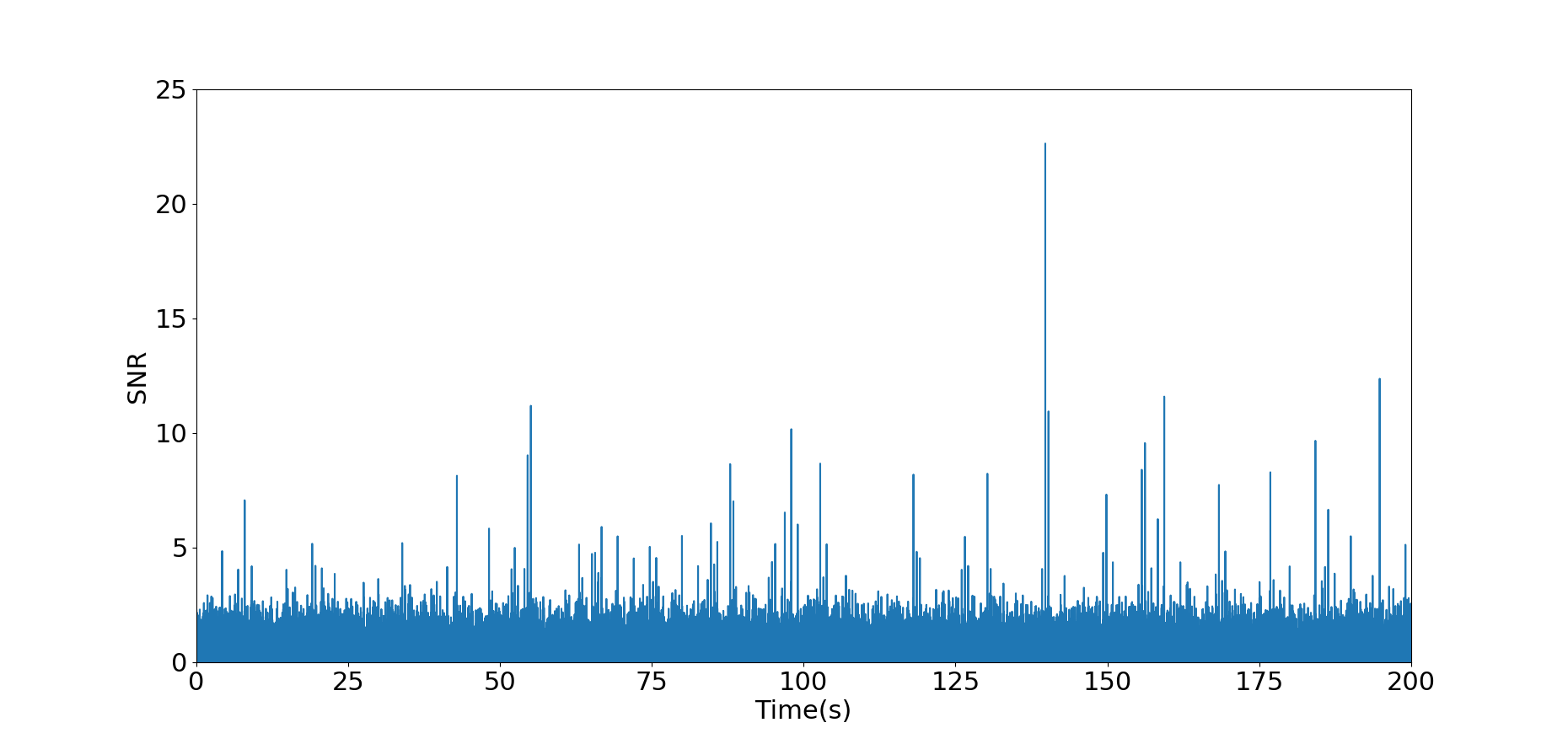}
        \includegraphics[width=0.8\textwidth]{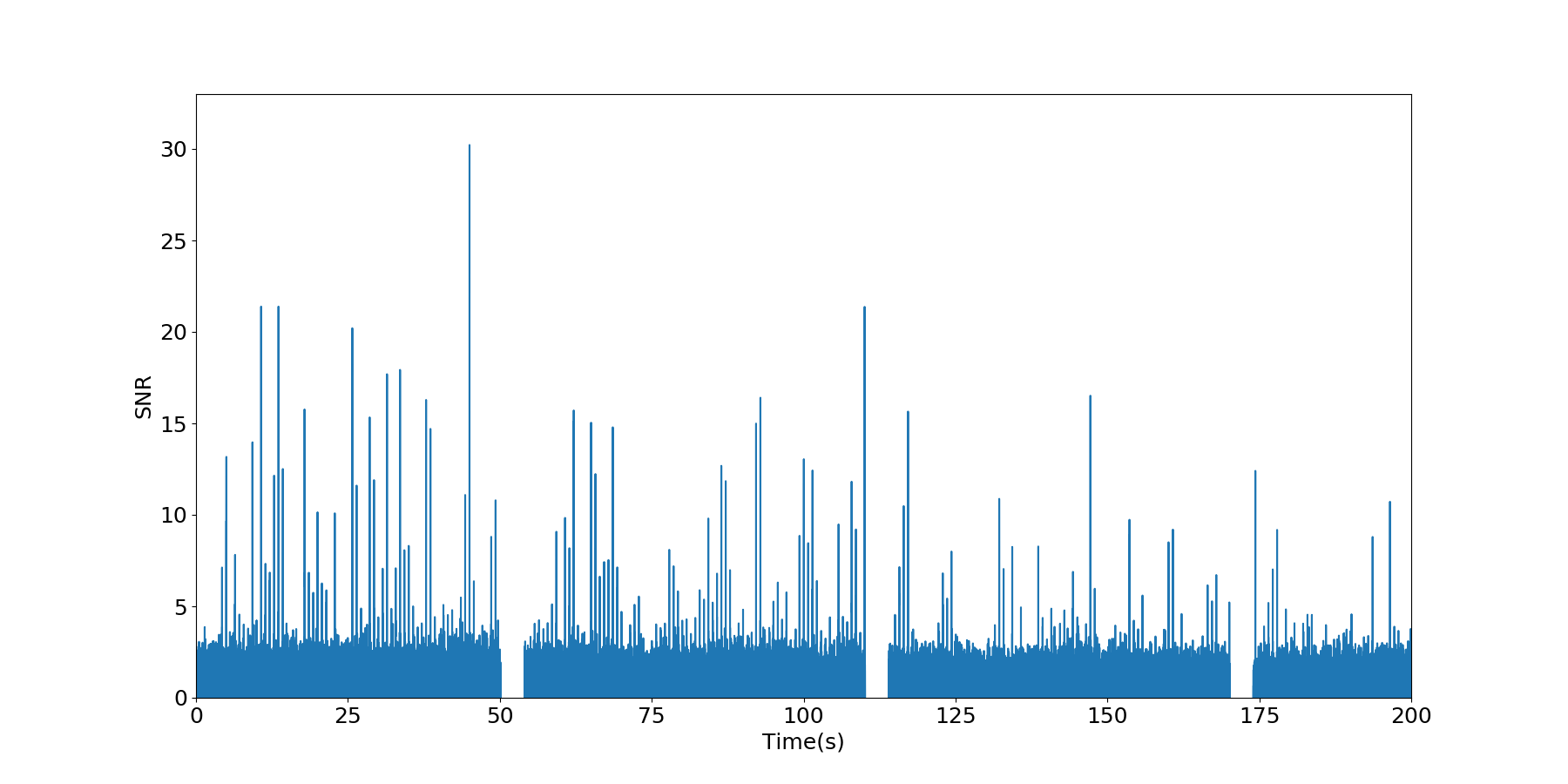}
    \caption{De-dispersed time series observing PSR B2021+51 (top) and B0329+54 (bottom).}
    \label{fig:B2021+51}
\end{figure}

We further validate the system by observing  PSR B0329+54 and PSR B2021+51. 
We succeed in detecting pulses from both pulsars, as shown in Fig. \ref{fig:B2021+51}. As our beams overlapped with adjacent ones, pulsars are visible in 2 or 3 beams simultaneously during observation.

\subsection{Sensitivity}
We can estimate the sensitivity of system by
using the data in Sec.\ref{sec:calibration} and Eq.(\ref{eq:sensitivity}), 
\begin{equation}
{\rm Sensitivity} =  \frac{S_{M1}}{I_{\rm peak}-I_{\rm noise}}*\Delta t*3\sigma_{\rm noise}
\label{eq:sensitivity}    
\end{equation}
our $3\sigma$ fluence sensitivity at the centre of primary beam is estimated to be 7.6 Jy ms.

\begin{figure}
    \centering
    \includegraphics[width=0.7\textwidth]{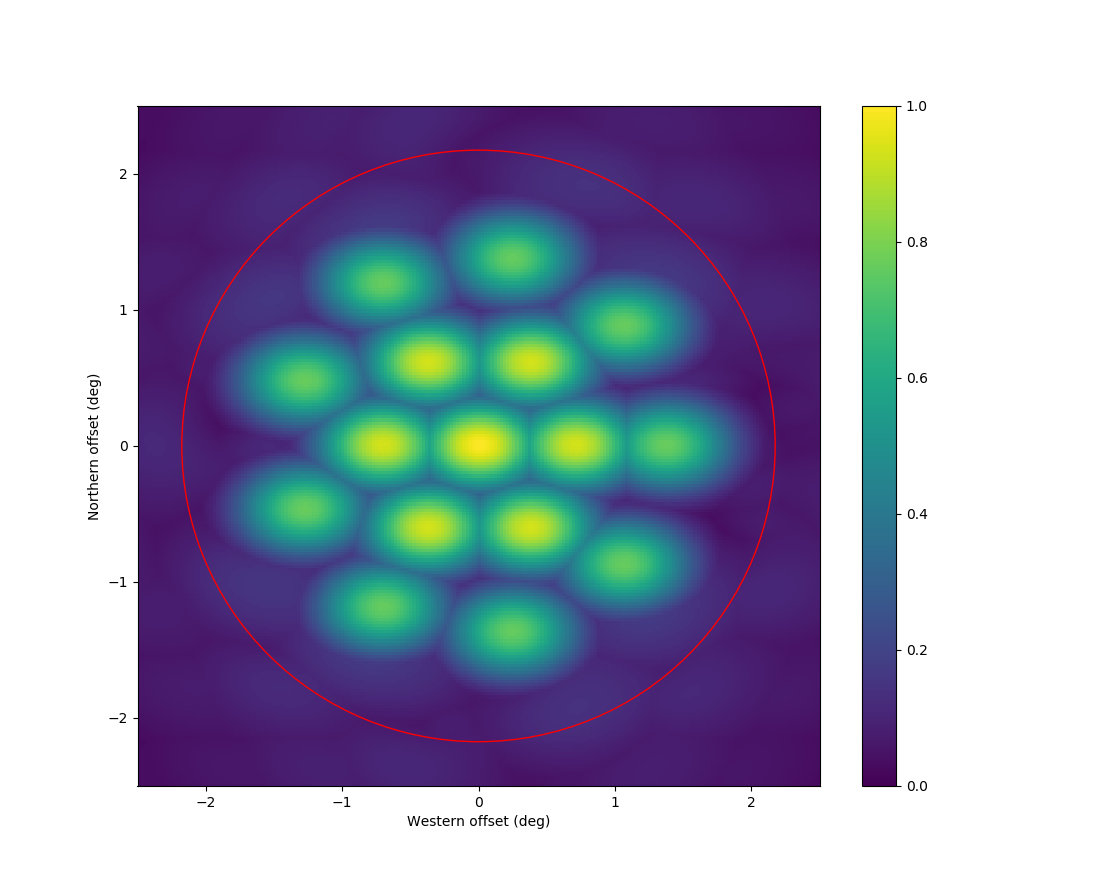}
    \caption{Overall beam profile. Red circle indicates the -3dB area of the primary beam.}
    \label{fig:beam_profile.png}
\end{figure}

We arrange our 16 digital beams in the following way: one is centered at the NCP during our regular survey, and the other 15 beams as two concentric rings around the center beam, one configuration (the optimized one) is shown in Fig.\ref{fig:beam_profile.png}. To maximize our FRB detection rate, we optimized three parameters: the number of beams in the inner circle, and the diameter of the inner and outer circles. We create a function that takes the above parameters as input and returns the estimated overall FRB detection rate. 

To obtain an estimated FRB detection rate for a given configuration, first we calculate the position of the beam centers, then obtain each beam's profile by shifting the digital beam profile to the given beam center, and multiplying it with the primary beam profile. For a given pixel, we estimate its effective detecting beam profile by using the beam with the best gain on that pixel, usually this is the beam with the nearest beam center. The FRB rate of every pixel is then calculated based on this overall beam profile. 

\begin{figure}
    \centering
    \includegraphics[width=0.7\textwidth]{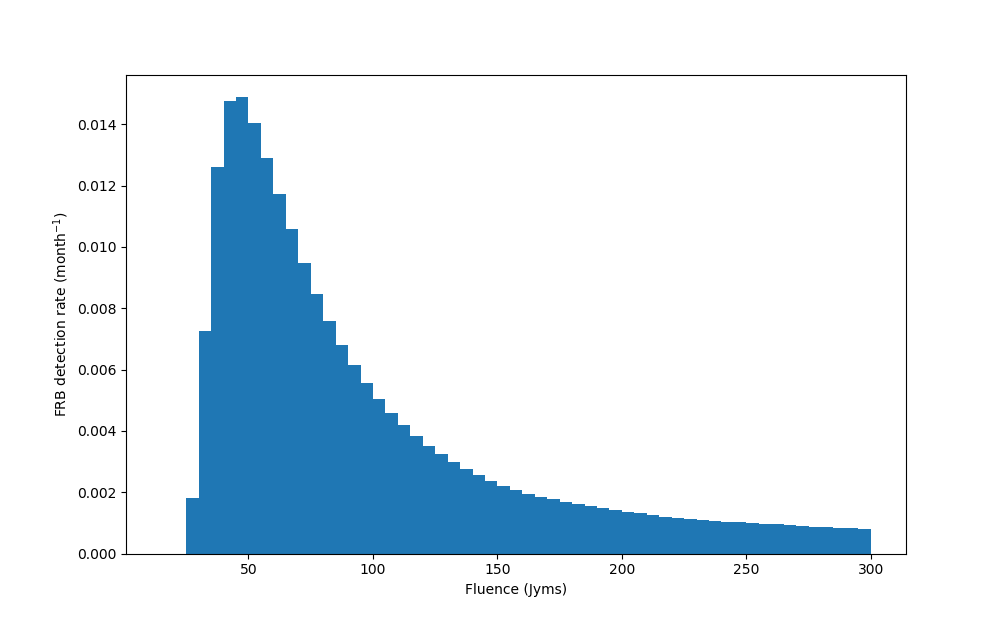}
    \caption{The expected differential detection rate of 10$\sigma$ FRBs as a function of fluence. Each bar indicates FRB rate within 5 Jyms. The FRBs with fluence greater than 300 Jyms are not shown in the figure but has a total rate of $\sim$ 0.05 $\rm month^{-1}$.}
    \label{fig:fluence_distribution}
\end{figure}

We take the all sky FRB rate as 820 per day above 5 Jyms at 400-800 MHz and fluence index $\alpha = -1.4$, as estimated from CHIME observations \citep{CHIME202112}. The overall FRB detection rate is obtained by summing over all pixels. Last, we find the maximum value of this function by varying the parameters. The optimal configuration with the above general lay out is to have 6 beams in the inner circle, and 9 beams in the outer circle, with diameters of  $1.46^\circ$ and $2.82^\circ$ respectively.  
The corresponding FRB rate is $\sim$ 1.2 FRBs per month for 3$\sigma$ bursts and $\sim$ 0.27 FRBs per month for 10$\sigma$ bursts, and the expected distribution of 10$\sigma$ FRB fluence is shown in Fig.\ref{fig:fluence_distribution}. 

Note that this is a rough estimate based on simplified treatment, because in reality, the signal-to-noise ratio depends not only on the fluence of the burst, but also on the shape of the signal, the distribution of the noise over the frequency band, and the RFI. We have also ignored the additional information provided by nearby beams, which could be important for the pixels near the border of adjacent beams.

\section{Conclusion}
\label{sec:Conclusion}

We have designed and installed the FRB-searching backend for the Tianlai dish pathfinder array, using a combination of FPGA beamformer, and an off-shelf GPU server for de-dispersion. The system can form 16 beams for each polarization. The system has been validated with test observation of pulsars.  The Tianlai Dish pathfinder Array will conduct a deep survey of the north celestial polar region by continuously pointing its telescopes to the NCP region\citep{Perdereau:2022ksl}, and the FRB search can be conducted concurrently.  
The data processing pipeline can make real-time processing of the data, and automatically raise alerts for possible follow up observation by other telescopes. 
The estimated $10\sigma$ sensitivity at the beam center is 25.3 Jy ms. For an FRB distribution estimated by \citet{CHIME202112}, a simple estimate gives an expected FRB detection rate of $\sim$ 0.27 FRBs per month. 

\begin{acknowledgements}
This work is supported by the Ministry of Science and Technology (MOST)
2018YFE0120800, National Key R\&D Program 2017YFA0402603, the National Natural Science Foundation of China (NSFC) grants 11633004 and 11473044, the Chinese Academy of Sciences (CAS) grants QYZDJ-SSW-SLH017.
The Tianlai Dish Array FRB backend is built with the support of the NAOC Special Fund for Preliminary Research. The Tianlai Dish Array was built with the support of the CAS special fund for repair and purchase. The Tianlai arrays are operated with the support of NAOC Astronomical Technology Center. 
\end{acknowledgements}

\bibliographystyle{raa}
\bibliography{refs}

\label{lastpage}

\end{document}